\documentclass{emulateapj}

\newcommand{\be}{\begin{equation}}
\newcommand{\ee}{\end{equation}}
\newcommand{\beqn}{\begin{eqnarray}}
\newcommand{\eeqn}{\end{eqnarray}}

\newcommand{\gn}{{{\cal G}}}
\newcommand{\pomega}{{{\varpi}}}
\def\yrs{\, \rm yrs}

\def\km{\, \rm km}

\def\cm{\, \rm cm}
\def\g{\rm g}

\begin{document}

\title{On the Origin of Pluto's Minor Moons, Nix and Hydra}
\author{Yoram Lithwick\altaffilmark{1} \& Yanqin Wu\altaffilmark{2}}
\altaffiltext{1}{CITA, Toronto ON Canada}
\altaffiltext{2}{Dept. of Astronomy \& Astrophysics, University of Toronto, Toronto ON Canada}

\begin{abstract}
How did Pluto's recently discovered minor moons form? Ward and Canup
propose an elegant solution in which Nix and Hydra formed in the
collision that produced Charon, then were caught into corotation resonances
with Charon, and finally were transported to their current location as
Charon migrated outwards.  We show with numerical integrations that,
if Charon's eccentricity is judiciously chosen, this scenario works
beautifully for either Nix or Hydra.  However, it cannot work for both
Nix and Hydra simultaneously.  To transport Nix, Charon's eccentricity
must satisfy $e_C< 0.024$; otherwise, the second order
Lindblad resonance at 4:1 overlaps with the corotation resonance,
leading to chaos.  To transport Hydra, $e_C > 0.7 R_{\rm
Pluto}/a_{\rm Charon}>0.04$; otherwise migration would be faster than
libration, and Hydra would slip out of resonance.  These two
restrictions conflict.  Having ruled out this scenario, we suggest an
alternative: that many small bodies were captured from the  
nebular disk, and they were responsible for forming, migrating and damping 
 Nix and Hydra.  If this is true,
small moons could be common around large Kuiper belt objects.
\end{abstract}

\section{Introduction}

The recent discovery of Pluto's two minor moons Nix and Hydra
\citep{Weaver,BGYYS06} presents interesting puzzles. We summarize
current observational data in Table \ref{table:parameters}.

\begin{table*}[t]
\caption{Observed parameters}
\begin{tabular}{ll | llll}
\hline\hline
&[units]\tablenotemark{a}&Pluto&Charon\tablenotemark{b}&Nix\tablenotemark{c}&Hydra\tablenotemark{c}
\\ \hline
  orbital period& [$T_C$] & 1 & 1 & 3.89 & 5.98 \\
  semi-major axis& [$R_P$] & 1.96 & 16.81  & 41.82 & 55.65 \\
 mass &[$M_P$] & 1 & 0.1165 & $7.8\times 10^{-6}$ & $1.8\times 10^{-5}$ \\
 eccentricity& & 0.0 & 0.0 & 0.002 & 0.005 \\
 inclination &[deg] & 96.14 & 96.14 & 96.18 & 96.36 \\ \hline
 \end{tabular}
 \label{table:parameters}
\tablenotetext{a}{Pluto's radius and mass  are $R_P=1164 \km$ \citep{Young},
and $M_P=1.3\times 10^{25} \g$.
Orbital 
period of Pluto-Charon binary is $T_C=6.3872$ days.
}
\tablenotetext{b}{Charon's radius and mass from {\citet{Sicardy05}. }}
\tablenotetext{c} {Nix and Hydra's radii and masses 
asssume a Charon-like albedo of $0.35$ \citep{Weaver} and a density of
$2\g/\cm^3$ (Charon-like); masses will be $\sim 20$ times higher (and
radii $\sim 2.7$ times larger) if albedo is $0.04$ (comet-like). 
Orbital parameters are from \citet{BGYYS06}.}
\vspace{0.5cm}
\end{table*}

\begin{itemize}

\item The orbits of the moons are nearly circular and  nearly coplanar
with Charon's orbit.

\item Nix, the inner minor moon, lies just inward of the 4:1 resonance with
Charon, while Hydra is close to the 6:1 resonance.

\end{itemize}

The Pluto-Charon system is doubly synchronized and circularized -- it
must have gone through significant tidal evolution in the past.  In
the currently favoured theory for the formation of Charon
\citep{Canup,McKinnon}, a giant impact chipped off a piece of the
proto-Pluto, leaving Charon on an eccentric orbit close to a rapidly
spinning Pluto.  Subsequent tidal evolution slowed down Pluto's spin,
pushed out Charon to its current position, and damped its orbital
eccentricity.  This is similar to how Earth's Moon is thought to have
formed and evolved.  Tides on Pluto pushed  Charon out to its current
position in $\sim 2\times 10^7 \left(Q_P/100\right)\yrs$, where $Q_P$
is Pluto's tidal quality factor.\footnote{ 
Estimates of various tidal timescales are  listed in Appendix B.
}
Charon's eccentricity
eventually decayed to zero on a comparable timescale, assuming that
Charon's tidal parameter $Q_C$ is not too different from $Q_P$.\footnote{Its
eccentricity would initially have grown if $Q_C/Q_P$ exceeds a number
that is of order unity; otherwise, its eccentricity would have
decreased monotonically.}  By contrast, Nix and Hydra likely cannot evolve
tidally at their current positions in the age of the Solar System
\citep{Stern,LWdamp}.

\section{Forced Resonant Migration (FRM)}
\subsection{The FRM Scenario}
\label{sec:frm}
\cite{WardCanup06} propose an elegant scenario to account for Nix
and Hydra's observed orbital properties.  In their scenario, the minor
moons were formed as byproducts of the collision that formed
Charon. Nix was then caught into Charon's 4:1 corotation resonance,
and Hydra into the 6:1 corotation resonance.  As Pluto's tides pushed
out Charon, Nix and Hydra remained in resonance and so they too were
forced to migrate towards their current orbits.  In this scenario, Nix
and Hydra must have been caught into the corotation resonance, and not
into any of the other sub-resonances at 4:1 and 6:1, because migration
in other sub-resonances would have excited the eccentricities of Nix
and Hydra to values much larger than are observed.

A corotation resonance can transport a particle only if Charon's
eccentricity $e_C$ is sufficiently large, because the resonant
libration time must be shorter than the time for the resonance to
migrate a distance of order its width, and the libration time
increases with decreasing $e_C$ whereas the width decreases.  The more
stringent constraint is set by Hydra, which requires $e_C\gtrsim
0.7(100/Q_P)^{1/5}(R_P/a_C)$, where $a_C$ is Charon's semimajor axis
\citep{WardCanup06}.  Therefore when Charon reached its current orbit
at $a_C\simeq 17 R_P$, it must have still had an eccentricity of
$e_C\gtrsim 0.04$.  As $e_C$ was subsequently damped by tides, the
width of the corotation resonances shrunk to zero, and Nix and Hydra
escaped from resonance.  Such a history for Charon's orbit is
plausible, given the uncertainties in the tidal parameters.

\cite{WardCanup06} briefly address the question of how the minor moons were 
initially trapped into corotation resonances.  If Nix and Hydra were
produced in a collision, then their free eccentricities were initially
large, and it would have been unlikely that they had just the right
orbits to end in corotation.  But if a lot of debris was produced in
the collision, and if this debris was highly collisional, then it is
possible that the debris settled into a cold disk with very little
free eccentricity,\footnote{The particles try to settle into the 
"coldest" orbits possible, with nested orbits that do not intersect each other if possible.}
 and that Nix and Hydra's free eccentricities were
damped by this disk. It is also possible that Nix and Hydra formed
from this debris.

We note that it is not possible that Nix and Hydra were formed outside
of corotation, and then were collisionlessly caught into corotation by
a migrating Charon.  The reason for this is that the contribution of a
corotation resonance to the Hamiltonian is a cosine term with constant
coefficient.\footnote{See, for example, the $c_0$ term in equation
(\ref{eq:hamfin}).  To be more precise, the coefficient of a corotation
resonance does not depend on the eccentricity of the particle, i.e.,
Nix or Hydra.  It does depend on the eccentricity of Charon, and on
the semimajor axis through the Laplace coefficient.  However, the
coefficient changes very little as the resonance sweeps over the
particle.}  By the symmetry of such a term, the energy lost when the
particle approaches the separatrix (the ``balance of energy'') must
equal the energy gained when the particle leaves the separatrix, and
capture is impossible \citep[e.g., equation 68 in][]{Henrard82}.

A potential difficulty with the FRM scenario not addressed by
\cite{WardCanup06} is the effect of the forced secular eccentricity.
If Nix and Hydra resided in corotation resonances when Charon's
eccentricity was high $(e_C\gtrsim 0.03)$, then their forced secular
eccentricity was comparable to $e_C$.  Yet their current eccentricity
is $\ll 0.03$.  
 How was this eccentricity damped?  The answer is that
their secular forced eccentricity tracks $e_C$, and that as $e_C$ decays, so
does the forced secular eccentricity---as long as the decay time of
$e_C$ is longer than the secular precession time.  To verify this, we
start from the secular part of the Hamiltonian for a test particle in
the presence of Charon (see Appendix \ref{sec:hamapp})
\be
H_{\rm sec}(Z) = -n\mu\left(e_CB_1{Z+Z^*\over 2} + B_2|Z|^2\right) \ ,
\ee
where $Z\equiv ee^{-i\pomega}$ is the particle's complex eccentricity, $n$ is its orbital frequency,
$\mu\equiv M_C/M_P$
and $B_1,B_2$ are sums of Laplace coefficients. The evolution of $Z$ is given by Hamilton's equation (eq. [\ref{eq:zham}]),
\beqn
{dZ\over dt}&=& 2i{\partial H_{\rm sec}\over \partial Z^*} \\
&=& -2in\mu \left(
e_C{B_1\over 2}+B_2 Z
\right) \ . \label{eq:sec}
\eeqn
The forced secular eccentricity is $e_{\rm forced}=-e_CB_1/2B_2$, and
the secular precession frequency is $\omega_{\rm sec}=2n\mu B_2$.  If
$e_C$ decays exponentially on timescale $\tau_e$, i.e.,
$e_C(t)=e_C\vert_{t=0}e^{-t/\tau_e}$, and if $Z\vert_{t=0}=e_{\rm
forced}$, then the solution of equation (\ref{eq:sec}) is
\be
Z=e_{\rm forced}\vert_{t=0}{ i\omega_{\rm sec}\tau_e e^{-t/\tau_e}-e^{-i\omega_{\rm sec}t}
\over i\omega_{\rm sec}\tau_e-1
}
\label{eq:forcee}
\ee
Therefore when $t\gg\tau_e$ and $e_C$ has decayed to $0$, the
particle's eccentricity will be negligibly small as long as
$\omega_{\rm sec}\tau_e\gg 1$, proving our assertion.

\subsection{Numerical Simulation of FRM}

\begin{figure}
\hspace{-.3cm}\vspace{-.5cm}
\includegraphics[width=9.cm]{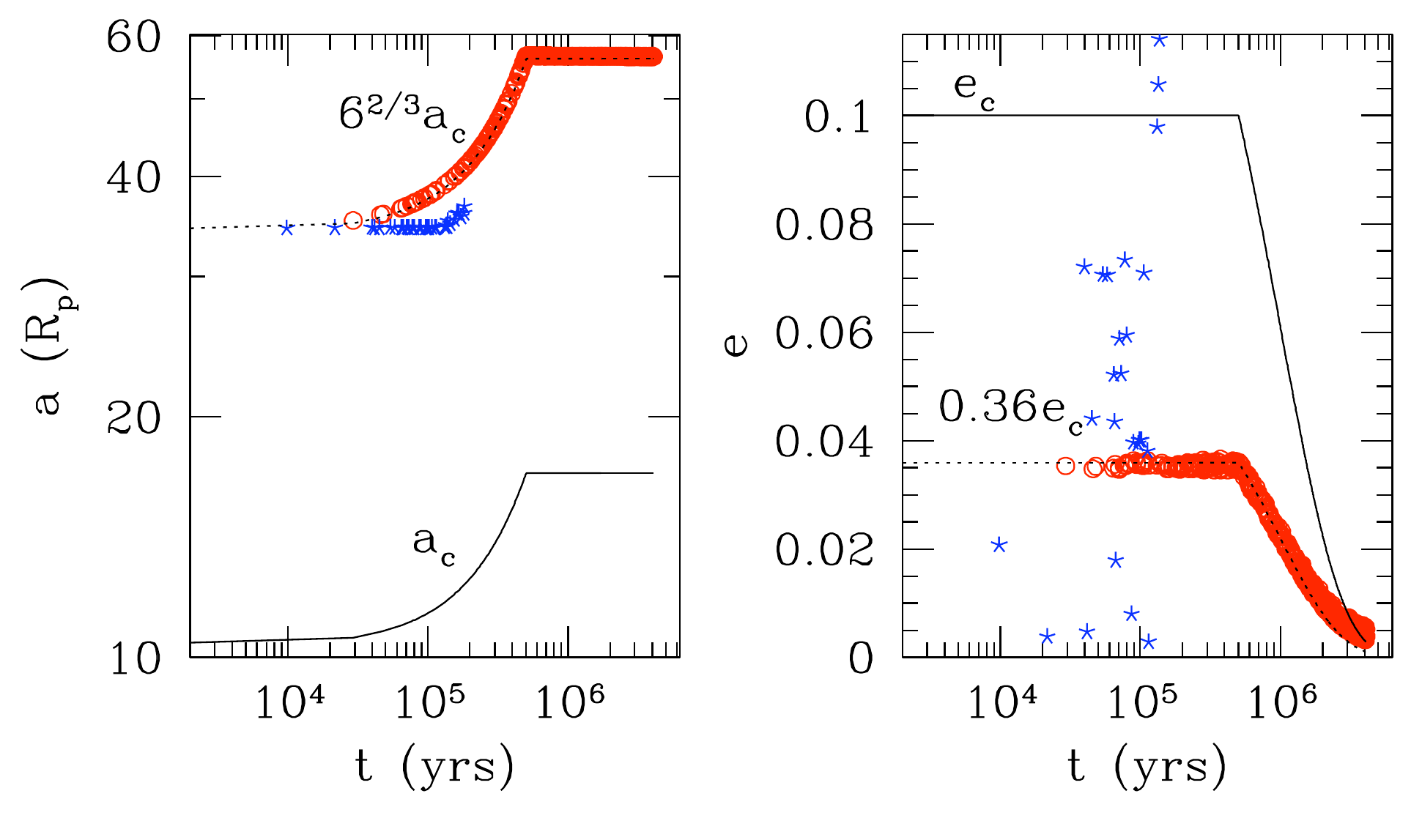}
\caption{
Numerical simulation illustrating the success of the \cite{WardCanup06} forced resonant migration scenario.
Charon is forced to follow a simple migration trajectory: it
 is first pushed
outward with a timescale of $10^6$ years from $10$ to $17$ pluto radii
(the left panel shows its semi-major axis as a function of time as a solid line), 
keeping its eccentricity fixed at $0.1$  (solid line in right panel).  After this we
force its eccentricity to decay in $10^6$yrs. A test particle initially placed
in Charon's 6:1 corotation resonance is resonantly migrated outward (dotted curves and
open circles). Notice that its eccentricity falls off 
following that of Charon. The value of its secular forced eccentricity is
$0.36 e_C$ at the 6:1 location.
A particle starting at the same location but having
a velocity that is larger by $0.1\%$ is quickly ejected
(star symbols).  To
reduce scatter, we have plotted the test particle values only when all
bodies have true anomaly $\lambda \sim 0$.
}
\label{fig:wcworks}
\end{figure}

Figure \ref{fig:wcworks} demonstrates numerically that the FRM scenario
works for Hydra when Charon's migration trajectory is chosen judiciously.
We start Charon  on an orbit with $e_C = 0.1$,
$a_C = 10 R_P$ and force it to migrate outward to $a_C = 17 R_P$ with $\dot
a = a/10^6$ yrs, keeping the eccentricity constant. Then $e_C$ is
forced to decay to zero in $10^6$ yrs.
A massless test particle representing Hydra is initially placed in the center
of Charon's 6:1 corotation resonance.  (See \S \ref{sec:upper} for a description of
how we start off a particle in the corotation resonance).  
 The orbital motions of all bodies are numerically integrated with the 
 SWIFT package \citep{LD94}, using the hierarchical Jacobi symplectic
 integrator of \cite{Beust}. We further modify it to allow semi-major axis and eccentricity
 evolution due to external forces, following the approach of \citet{Leetide}.

Figure \ref{fig:wcworks} shows that Charon indeed pushes out Hydra. Hydra remains
within the corotation island until Charon's eccentricity falls to a very small value. Hydra's
eccentricity tracks that of Charon's along the way (eq. [\ref{eq:forcee}]).
Fig. \ref{fig:wcworks} also shows that FRM depends
extremely sensitively on the initial
conditions of the test particle. A second particle started off at exactly the
same location as the one above but with a velocity larger by $0.1\%$ (and therefore
falling outside the narrow corotation island) is quickly ejected during the migration of Charon.
In fact, the width of the corotation island is so narrow that it is comparable to the size of
Hydra at the start of  our experiment. 

\section{Ruling out FRM}

 FRM can migrate either  Hydra or Nix to their current
orbits.
But it cannot simultaneously migrate both of them.  To transport
Hydra, Charon's eccentricity must be greater than a critical
value; otherwise, Hydra would slip out of resonance.  To transport
Nix, Charon's eccentricity must be less than a second critical value;
otherwise, the 4:1 corotation resonance would be destroyed by
resonance overlap.  These two constraints conflict.  We discuss them
in turn.

\subsection{Lower Limit on $e_C$ from Hydra's migration}
\label{subsec:lower}

To be able to transport Hydra in resonance, the migration time across
the width of the 6:1 corotation resonance must be longer than the
libration period in the resonance,
\be
6^{2/3}\dot{a}_C\lesssim {\Delta a_{\rm lib}\over T_{\rm lib}} \ ,
\ee
where $\Delta a_{\rm lib}$ is the resonance width, $T_{\rm lib}$ is
the libration period, and $\dot{a}_C$ is Charon's tidal migration
rate.  

\begin{figure}
\hspace{-.2cm}\vspace{-2cm}
\includegraphics*[width=9.3cm]{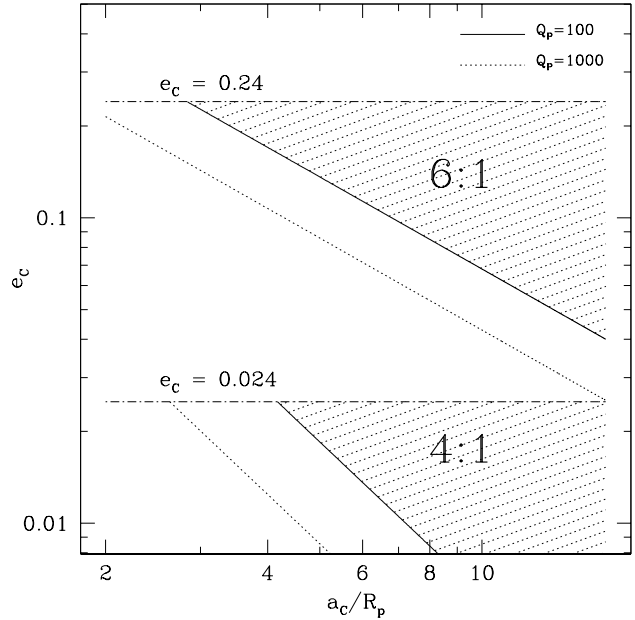}
\caption{This figure shows why FRM cannot work. The solid  lines show the lower 
limit of $e_C$ below which
tidal migration of Charon is too rapid for forced resonant migration for $Q_P=100$.
These lines are
given by eq. [\ref{eq:ecmin}] for the 6:1 resonance, and an analogous
expression for the 4:1 resonance.
 The dotted lines show the same limits, but for $Q_P=1000$.
 The dot-dashed horizontal 
lines show the maximum values of $e_C$, above which individual corotation
resonances are destroyed.  For the 4:1 resonance
one must have $e_C<0.024$.
FRM is feasible for each resonance only if $e_C$ lies between these two limits
(shaded region, for $Q_P=100$).  The lower limit for the 6:1 resonance lies
above the upper limit for 4:1 resonance. Hence Charon could not resonantly
migrate Nix and Hydra simultaneously.
}
\label{fig:wcnotwork}
\end{figure}

We take the expressions for the libration width and libration period from
equations (8.58) and (8.47) of \citet{MD99},  and use the Kaula formula 
\citep[\S 6.3 of][]{MD99} to obtain
a resonance strength for the 6:1 corotation resonance  of $f_d  \approx
0.02$ \citep[as defined in {eq. [8.32]} of][]{MD99}. Adopting the tidal timescale for
orbital expansion as in  equation \ref{eq:time2}, the above transport condition
translates to a lower limit for $e_C$,
\be
e_C> 0.04 \, \left( {{k_{2P}}\over 0.05} \right)^{1/5}\,
 \left( {{100}\over Q_P} \right)^{1/5}\,
\left({{17 R_P}\over a_C}\right) \ . \label{eq:ecmin}
\ee
Note the weak dependence on $k_{2P}$ (Pluto's tidal Love number) and
$Q_P$ (Pluto's tidal quality factor).
The above limit has also been given in \cite{WardCanup06}.  We have performed
numerical experiments to confirm the numerical coefficient.
 We plot  this limit  in Figure
\ref{fig:wcnotwork}.

\subsection{Upper Limit on $e_C$ from Resonance Overlap At Nix's Orbit}
\label{sec:upper}

In this subsection, we demonstrate with two numerical experiments
 that Charon's 4:1 corotation resonance exists only if $e_C<0.024$. 
Otherwise, resonance overlap destroys the corotation resonance 
(as shown in \S \ref{sec:destroy}).
Figure \ref{fig:wcnotwork} shows that the constraint $e_C<0.024$, together
with the constraint from \S \ref{subsec:lower}, rules out the FRM scenario:
it is impossible to satisfy both of these constraints simultaneously.

To demonstrate numerically the destruction of the 4:1 corotation resonance, one
first needs a robust method to find it when it exists.  
This is not entirely trivial because the mass ratio $M_C/M_P$ is not terribly small,
and neither are Charon's and Nix's eccentricities $e_C$ and $e$. 
Hence an expansion of the disturbing function will not be very accurate.
Instead, our method is based
on finding the ``coldest'' orbits of test particles around the Pluto-Charon 
binary.  A disk of infinitesimal particles that collide inelastically will naturally
tend to the coldest orbits, i.e. the orbits that minimize the velocity dispersion
within the disk.
What are these coldest orbits?  Consider first the case that Pluto and Charon
orbit each other on circular orbits.  
Then the coldest orbits are the periodic orbits, which
are orbits that, in the rotating reference frame of the binary,
close on themselves after a single loop around the binary. 
The generalization of periodic orbits to the case of non-zero $e_C$
are invariant loops \citep{MS97,PSA05}, which may be understood
as follows.
If we take a snapshot of the position of an orbiting particle every
time Pluto-Charon reach some predefined orbital phase (periapse, say),
then in general the snapshots would fill out a two-dimensional region
in the $r,\theta$ plane, where $r,\theta$ are the particle's radius
and azimuth.  But the coldest orbits---the invariant loops---are those
in which the snapshots trace out a one-dimensional closed loop.
Far from strong
resonances, neighbouring invariant loops do not intersect each other,
or themselves.  But near strong resonances, they often do intersect.

 To find the corotation resonance, we find the set of invariant
loops around an eccentric Pluto-Charon binary, largely following the
procedure of \cite{PSA05}.  We express the orbit of the binary in
terms of the eccentric anomaly, expanded to fourth order in $e_C$.
The motion of the test particle is evolved in the barycentric frame
with a 4th-order Runge-Kutta integrator with adaptive step-sizes
\citep{Pressetal}, with an error tolerance of $10^{-8}$.  A test
particle is initially launched far from any strong resonances on the
periastron axis ($x$-axis) when Charon is at periapse.  We assume that
its orbit is symmetric with respect to the $x$-axis, so the only
unknown is the velocity $v_y$.  This is initially guessed using the
local Keplerian value.  The positions of the test particle at each
subsequent periapse passage of Charon are recorded, for a total of
$10^4$ binary periods.  These are separated into $2,000$ angular bins
and the radials dispersion within each angular bin is co-added.\footnote{This
is different from \cite{PSA05} in that they only use the dispersion
near the $x$-axis.  We find that our treatment gives a faster and more
reliable convergence.}  We use the bisection technique (typically 
within a range 6\% of the initial guess) to find the correct $v_y$
that minimizes this dispersion. The resulting 1-D curve $r(\theta)$ is
an invariant loop.  We then proceed to find the next loop which is
closer to the resonance, using the previously obtained $v_y$ as the
initial guess.  The closer the loops, the better the guess, and the
more reliable the convergence.

\begin{figure}
\hspace{-.3cm}\vspace{-2cm}
\includegraphics[width=9.5cm]{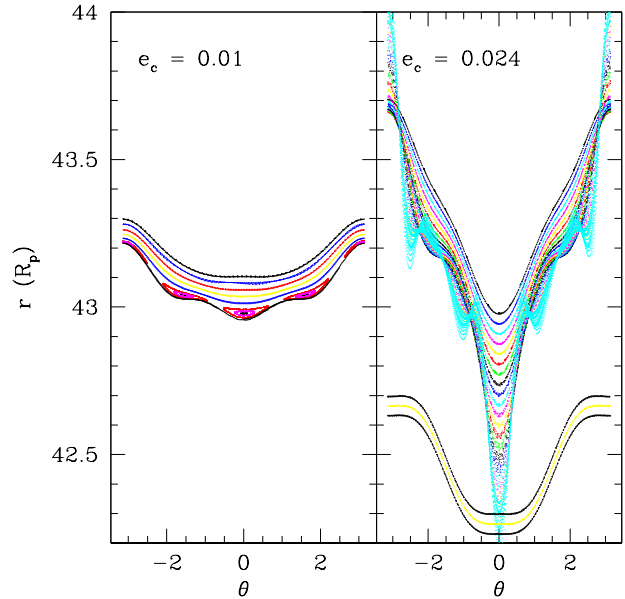}
\caption{Invariant loops
near the 4:1 location, showing the 4:1 corotation 
resonance (as 4 distinct islands) when $e_C = 0.01$ (left) and its
absence when $e_c = 0.024$ (right).  These loops represent 
orbits that are the ``coldest'' possible (no free eccentricity). 
Here, we record particle
positions (barycentric radius in unit of $R_P$, position angle $\theta$ 
as measured from Charon's periapsis) whenever Charon is at periapsis. 
The resulting surface-of-section is either a continuous line (no resonance)
or disjoint islands (in resonance).  For $e_C = 0.01$,
other  $4:1$ subresonances lie beneath the corotation resonance (not shown here); 
when $e_C = 0.024$,
subresonances overlap and there are no periodic orbits within the resonant region.
}
\vspace{.2cm}
\label{fig:xx}
\end{figure}

Fig. \ref{fig:xx} depicts the invariant loops we obtain for a test
particle near the 4:1 location. When $e_C < 0.024$, 4:1 corotation
islands are clearly visible; while when $e_C = 0.024$ and beyond, all islands
disappear.

\begin{figure}
\hspace{-.2cm}\vspace{-2cm}
\includegraphics[width=9.3cm]{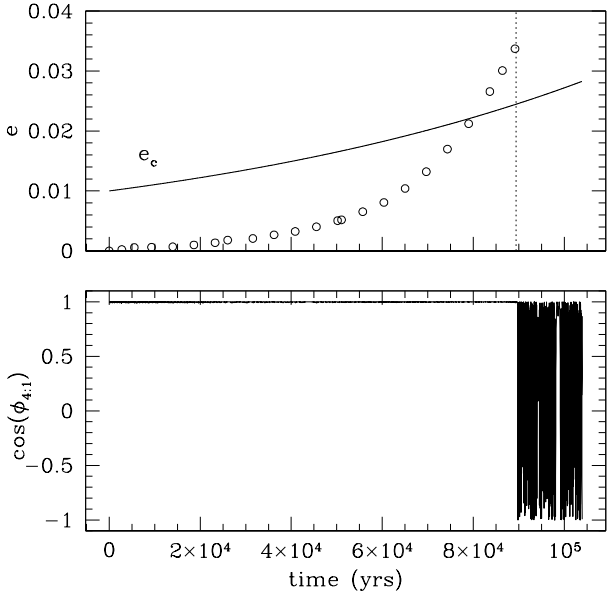}
\caption{Maximum $e_C$ above which the 4:1 corotation resonance disappears.
We place Charon at its current position, and a test particle at 4:1
corotation resonance.  As $e_C$ is increased gradually (solid curve in top panel,
$\dot e/e = 1/10^5$ years), the barycentric eccentricity of the
test particle rises even faster, until  $e_C = 0.024$ (marked by the dotted line) 
after which resonance overlap destroys the corotation resonance -- 
the resonant angle $\phi_{4:1} = 4\lambda - \lambda_C - 3 \tilde
\omega_C$ suddenly starts circulating. The particle is released from 
the resonance and displays chaotic motion. To reduce scatter, we only show test
particle data when both bodies have true anomalies $\lambda \sim 0$.
The same break-down for the 6:1 resonance occurs at $e_C = 0.24$.
}
\vspace{.5cm}
\label{fig:xxx}
\end{figure}

Fig. \ref{fig:xxx} shows the results from a second experiment that demonstrates
the destruction of the 4:1 corotation resonance for large $e_C$.
In this experiment, we insert a test
particle into the center of the 4:1 corotation resonance when $e_C$ is small, 
and then slowly raise
the value of $e_C$. As $e_C$ rises above $0.024$, the motion of a particle initially
trapped inside the 4:1 corotation resonance becomes
chaotic and the resonance angle circulates. 

Similar experiments performed for the 6:1 corotation find the
limiting $e_C$ to be $0.24$ in that case.

\section{Resonance Overlap Destroys the 4:1 Corotation Resonance}
\label{sec:destroy}

We seek a better understanding of how and why the 4:1 corotation
resonance is destroyed when $e_C>0.024$.  We model the evolution of a
test particle near the 4:1 resonance of an eccentric Charon-Pluto
binary with a truncated disturbing function, keeping only secular and
4:1 resonant terms.  The particle's equations of motion are compactly encoded by
the Hamiltonian derived in Appendix A (eq. [\ref{eq:hamfin}]), 
which has two degrees of freedom.  The momentum 
and co-ordinate of the first degree of freedom $\{p_a,q_a\}$ are related
to the test particle's semimajor axis and mean longitude via equations (\ref{eq:parelate})
 and
(\ref{eq:qarelate}).  And the momentum and co-ordinate of the second degree 
of freedom are combined into the complex canonical variable $z$, which
is the test particle's free complex eccentricity (i.e.,  the complex eccentricity
after subtracting the forced secular eccentricity, eqs. [\ref{eq:zdef}] and [\ref{eq:ztrans}]).
The equations of motion for these two degrees of freedom are just Hamilton's equations
(eq. [\ref{eq:eom13}]), which we shall numerically integrate.
 Since this is a
time-independent Hamiltonian with two degrees of freedom, phase space
may be mapped out with Poincar\'e surfaces of section 
\citep[e.g.,][]{HH64}

\begin{figure}
\hspace{-.2cm}\vspace{-2cm}
\includegraphics[width=9.3cm]{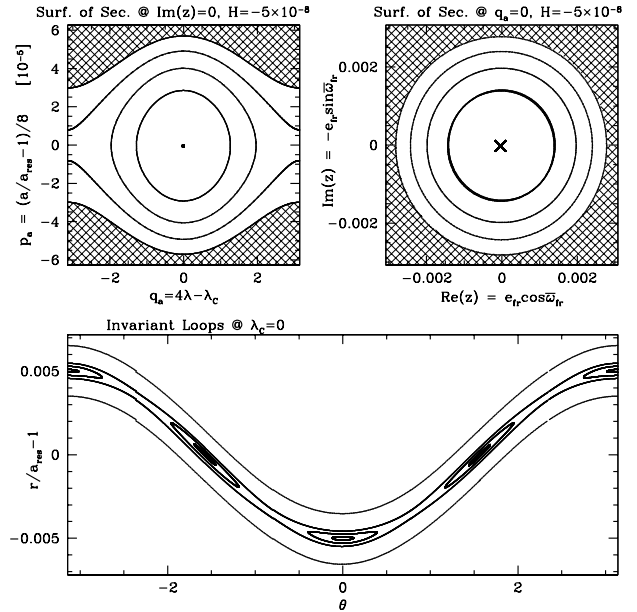}
\caption{
Orbits arising from the corotation resonance only, from numerical
integrations of Hamiltonian (\ref{eq:corot})'s equations of motion
 with $e_C =0.01$. Top two panels show
surfaces of section at a fixed value of the energy.  Cross-hatched
regions are incompatible with the chosen
energy and are forbidden.  
Top left panel shows values of $q_a,p_a$, recorded when Im$(z)=0$, for
seven orbits; top right panel shows values of $z$ when $q_a=0$ for the
same seven orbits (with two of the orbits overlapping two others, making
it appear as if only five orbits are plotted).
The fixed point in the $z$-plane marked by an `x'
is an invariant loop.  When this orbit's radius is plotted versus
its azimuth via equations (\ref{eq:tt})-(\ref{eq:rt}), it gives a 1D curve, 
as shown in the bottom panel.
 (Specifically, it gives the curve in the bottom panel that lies immediately
above the librating regions.)
Also shown in the bottom panel are five other invariant loops for five other energy values.
Compare this plot with Fig. \ref{fig:xx} obtained from full numerical integrations.
}
\label{fig:corot}
\end{figure}

We first consider  the corotation term (the $c_0$ term
in Hamiltonian \ref{eq:hamfin})  in isolation, 
discarding the Lindblad terms (the $c_1$-$c_3$ terms), which leaves
\be
H=-24p_a^2-\mu c_0e_C^3\cos q_a-\mu B_2 |z|^2 \label{eq:corot}
\ee
with the two degrees of freedom decoupled from each other;
 $c_0$ and  $B_2$ are order-unity constants whose values are listed
 in Appendix A.
  Hamilton's equations for
$q_a,p_a$ yield $\dot{q_a}=-48 p_a$ and $\dot{p_a}=-\mu c_0 e_C^3 \sin
q_a$, the same as for a pendulum.  Hence the
angle $q_a$ will either librate around the center of the resonance
(where $q_a=p_a=0$), or, for large enough $|p_a|$, it will circulate
(Fig. \ref{fig:corot}, top left panel). Hamilton's equation for $z$ is
$\dot{z}=-2i\mu B_2 z$, showing that the free eccentricity has
constant amplitude and a circulating phase (Fig \ref{fig:corot}, top
right panel). 

As in \S \ref{sec:upper}, we seek the ``coldest'' orbits, i.e. the
invariant loops.  
An invariant loop appears as a fixed point in the $z$-plane
surface of section.
 To see this, recall that an invariant loop is a 1D curve of the test
 particle's $r(\theta)$ (radius vs. azimuth) whenever Charon reaches periapse.
When Charon reaches peripase, the test particle has\footnote{
More generally, we should write $\theta=q_a/4+j\pi/2$, where $j$ is an integer.
The four corotation islands in the $r$-$\theta$ plane are produced by different values of $j$.
}
\beqn
\theta(t)&=&{q_a(t)\over 4} \label{eq:tt} \\
{r(t)\over a_{\rm res}}&=&1+8p_a(t)-{\rm Re}
\left(
e^{i\theta(t)}\left(
z(t)+e_{\rm forced}
\right)
\right) \label{eq:rt} \ ,
\eeqn
valid to first order in $z$ and $p_a$, where $a_{\rm res}$ and $e_{\rm forced}$
are constants defined in the Appendix.  Since the $z$-plane surface of
section has $\theta=q_a=0$, we see that an invariant loop must be
a fixed point in that section;  otherwise, there would be an infinite number
of different values of $z$ (and hence of $r$) at $\theta=0$, and the orbit would
not trace out a 1D curve in the $r$-$\theta$ plane.
In summary, to find the invariant loops, we first find the fixed point
in the $z$-plane surface of section.  We then plot the $r$ vs. $\theta$ for
this orbit, as given by equations (\ref{eq:tt})-(\ref{eq:rt}).
For example, the fixed point marked with an `x' in the upper-right panel of Figure \ref{fig:corot}
gives rise to one of the curves in the lower panel of that figure. The five other invariant loops
shown there
were found similarly, but with different values of the energy.

\begin{figure}
\hspace{-.2cm}
\includegraphics[width=9.cm]{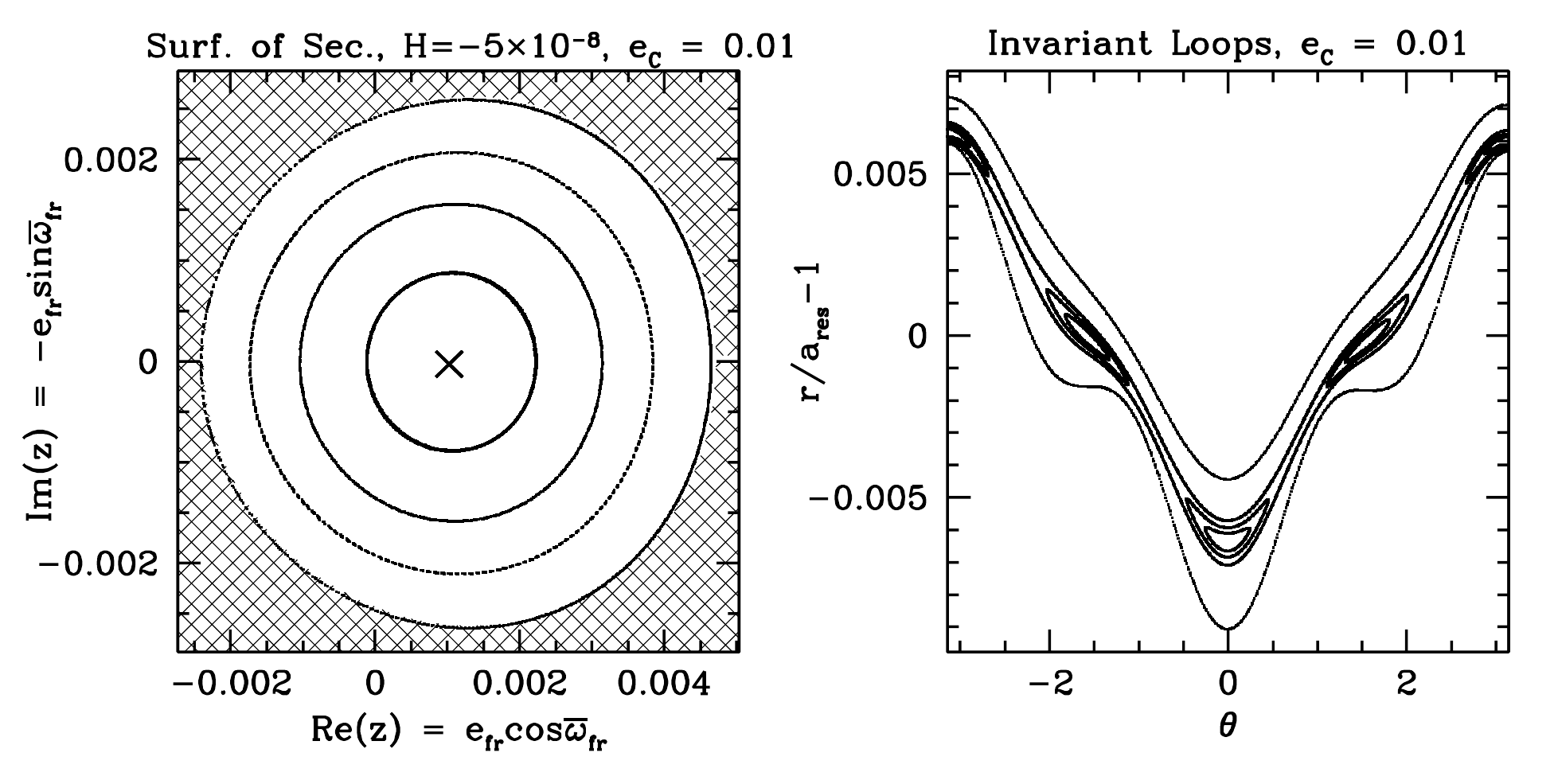}
\caption{
Surface of section (left panel) and invariant loops (right panel) from integration of the 
full Hamiltonian (eq. [\ref{eq:hamfin}]), when
$e_C=0.01$.  The fixed point in the surface of
section that is marked with an `x' produces the invariant loop that is
just above the librating retions.  Notice that
the inclusion of the Lindblad terms has shifted the value of $z$ for the fixed point
(eq. [\ref{eq:fo}]).} 
\label{fig:ecpt01}
\end{figure}

\begin{figure}
\hspace{-.4cm}\vspace{0cm}
\includegraphics[width=9.cm]{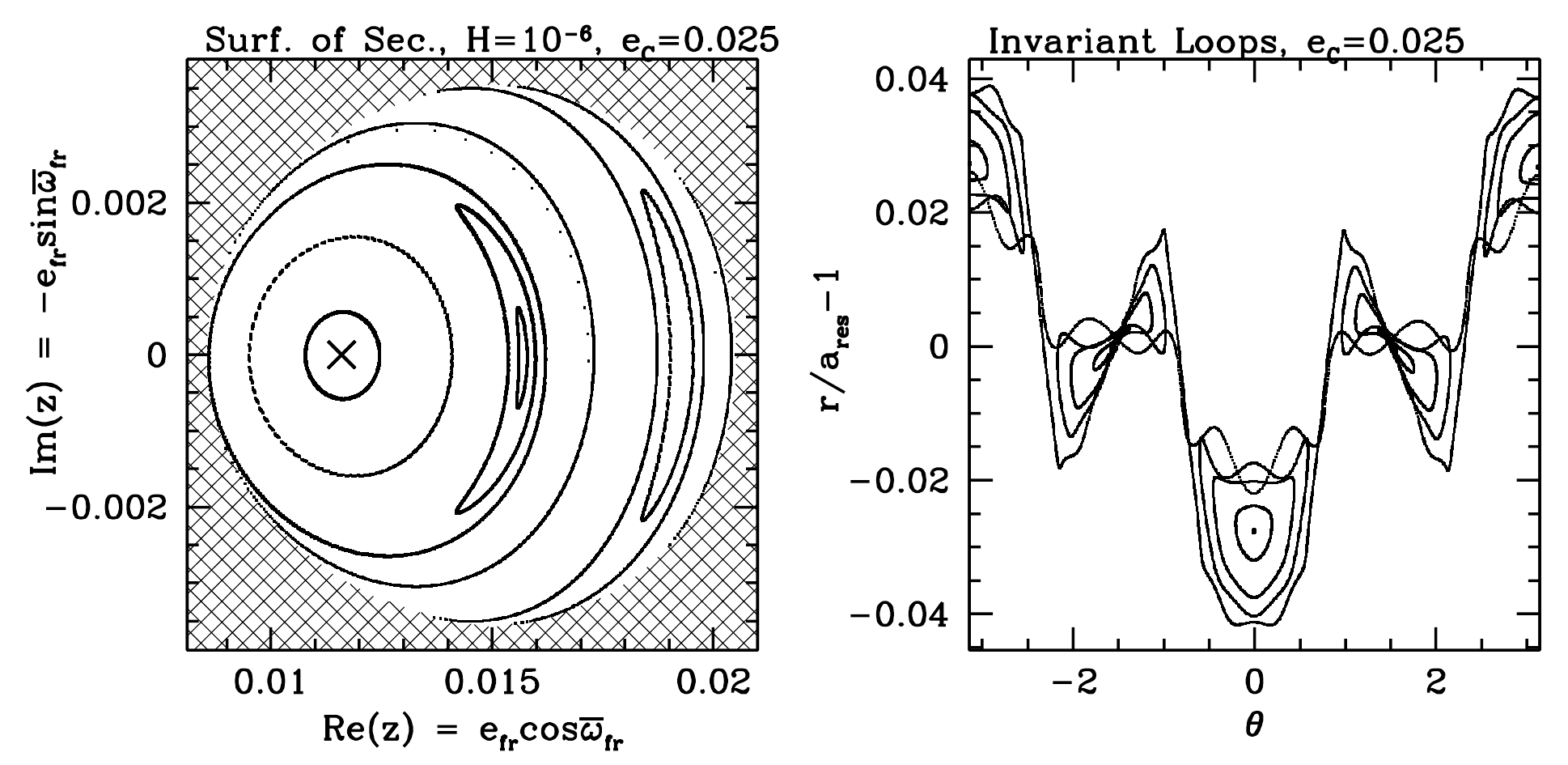}
\caption{Same as Fig. \ref{fig:ecpt01} but for
$e_C=0.025$.  The fixed point in the surface of section
marked with an `x' produces one of the librating invariant loops.  New
fixed points have appeared in the surface of section; these give
invariant loops that self-intersect (not shown). }
\label{fig:ecpt025}
\end{figure}

\begin{figure}
\hspace{-.2cm}\vspace{-2cm}
\includegraphics[width=9cm]{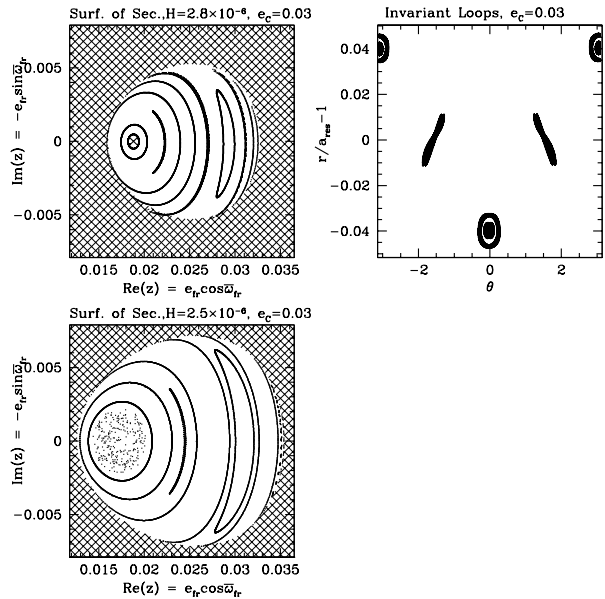}
\caption{Same as Fig. \ref{fig:ecpt01} but for
$e_C=0.03$: Two left panels show surfaces of section at two energies.
The fixed point (marked `x') in the upper-left panel produces the
outermost librating orbit shown in the plot of invariant loops.  When
the energy is decreased by a small amount to $2.5\times 10^{-6}$,
instead of producing an invariant loop with larger libration, the
fixed point disappears into a sea of chaos (lower-left panel).  The
chaotic orbit gives rise not only to the splattering of points shown
near $z=0.017$; its surface of section also has three chaotic islands
at large $|z|$.  These islands are not seen in the figure, because
they are off the scale.  }
\label{fig:ecpt03}
\end{figure}

Having described how to find the invariant loops, we turn now to the full
4:1 Hamiltonian (eq. [\ref{eq:hamfin}]).  Figure \ref{fig:ecpt01} shows
orbits when $e_C=0.01$. In the sample surface of section shown, the
fixed point is no longer at $z=0$ because the Lindblad resonances give
a non-zero forced eccentricity that is in addition to the forced
secular value.  The invariant loops in the right panel of the figure
are similar to those with only the corotation resonance
(Fig. \ref{fig:corot}), but are slightly more distorted.  When
Charon's eccentricity is increased to $e_C=0.025$, the invariant loops
become highly distorted (Figure \ref{fig:ecpt025}).  By $e_C=0.03$,
many of the librating invariant loops no longer exist---the fixed
points break up into a sea of chaos (Fig. \ref{fig:ecpt03}). At even
higher $e_C$, there are no corotation librations left.

In \S \ref{sec:upper} we found from direct integration of Newton's
equations that the corotation islands disappear when $e_C>0.024$.
Although the critical value we find in the present section $\sim 0.03$
is similar, we speculate that the reason for the discrepancy is that
our numerical coefficients are not very precise.  For example, we have
neglected terms that are $O(M_C/M_P)\sim 0.1$.

Why does chaos appear in the Hamiltonian model of equation
(\ref{eq:hamfin}) when $e_C\gtrsim 0.03$?  Chaos often appears when
separatrices of two neighbouring resonances overlap \citep{Chirikov79}.
However, it is difficult to apply this resonance overlap criterion to
our 4:1 Hamiltonian, because the separatrices have non-trivial shapes
in four-dimensional phase space.  Instead, we give two
semi-quantitative explanations.

First, we calculate the eccentricity that the first two Lindblad terms
($c_1,c_2$) force on a particle at the center of corotation.  The
equation of motion for $z$ at $q_a=0$ (center of the corotation resonance) is
\be
\dot{z}=-2i\mu\left(B_2 z + c_1 e_C^2+c_2e_C {\rm Re}(z)\right) \ ,
\ee
setting  $c_3=0$.  The forced eccentricity
is determined by setting $\dot{z}=0$, which yields
\be
{\rm Re}(z_{\rm forced})=-{c_1 e_C^2\over B_2+c_2 e_C} \ . \label{eq:fo}
\ee
The forced eccentricity is infinite when $e_C$ is given by
\be
e_{C*}\equiv -{B_2\over c_2}= 0.031
\ee
We note that \cite{WardCanup06} perform a calculation similar to our
equation (\ref{eq:fo}) in their Supplementary Notes, although they
include only the first Lindblad term (i.e., they effectively set $c_2=c_3=0$).  
We also note that if we do not set
$c_3=0$, the forced eccentricity no longer diverges, although it still
becomes large when $e_C\sim e_{C*}$.

For our second explanation, we argue that chaos is likely to occur
when the center of the second Lindblad resonance (the $c_2$ term)
overlaps the center of the corotation resonance in the $q_a,p_a$
plane. To find the center of the second Lindblad resonance, we switch
from $z$ to the canonical variables $q_e,p_e$, defined via
$\sqrt{2p_e}e^{iq_e}\equiv z$.
Then,  the Hamiltonian for the second Lindblad resonance becomes
$H=-24p_a^2-2p_e\mu B_2-2\mu c_2 e_Cp_e\cos(q_a+2q_e)$.  The center of
this resonance occurs where $q_a+2{q}_e=0$.  Employing Hamilton's
equations, we find $0=\dot{q}_a+2\dot{q}_e=-48 p_a-4\mu B_2-4\mu c_2
e_C$.  Therefore the center of this resonance overlaps the center of
the corotation resonance (which is at $p_a=0$) when $e_C$ is given by
$e_{C*}$.

\section{Alternative Formation Scenarios}

How did Nix and Hydra form?
\cite{WardCanup06} argue that these moons are byproducts
of the impact that formed Charon. 
But here we argue that this is not the case.
If Nix and Hydra were byproducts of the impact, one might
imagine three possibilities for how they ended up
in their current orbits, with very small eccentricities:
\begin{itemize}

\item They might have been directly ejected to their
current semimajor axes with large eccentricities,
which were subsequently
damped by tides.
However, the timescale
for them to damp their eccentricities by tidal
interaction with Pluto 
 is longer than
the age of the Solar System \citep{Stern}.\footnote{Nix could have
damped its eccentricity if it is a strengthless rubble pile
and has both tidal Love number and tidal Q factor of order unity;
but Hydra could not have, even in this extreme case.}
In addition, they likely could not have damped their
eccentricity by exciting Charon's eccentricity, with
Charon in turn tidally damping its own eccentricity \citep{LWdamp}.

\item  They might have been ejected from the Charon-producing
impact to their current
semimajor axes along with many small particles. 
If these small particles formed into a collisional disk, the disk could have damped
Nix and Hydra's eccentricities.  However, 
a post-impact collisional disk probably could not have extended to
such a large distance ($\sim 50 R_P$)  from Pluto: simulations of the Pluto-Charon
impact
give much more compact disks \citep{Canup}.

\item As proposed by \cite{WardCanup06}, they might have been 
damped by a collisional disk much closer to Pluto, and then been migrated outward
by Charon.  However, we have shown in this paper that 
 both Nix and Hydra could not
 be resonantly migrated outward by Charon in the 4:1 and 6:1 resonances, respectively.
Is it possible that Charon was responsible for resonantly migrating
only Nix ($e_C < 0.024$), while Hydra was transported outward in Nix's
3:2 corotation resonance?  We have failed to find such an orbit
using our algorithm (\S \ref{sec:upper}) in the presence of
an eccentric Charon and a massive Nix. But we have yet to exclude this possibility
with more confidence.

\end{itemize}

Taken together, the above results suggest that Nix and Hydra 
are not byproducts of the Charon-forming impact. In addition, 
Nix and Hydra could not have been captured from the Kuiper
belt, and subsequently collisionlessly hardened by other passing-by
bodies.  Although that mechanism might work for other Kuiper belt binaries
\citep{GLS02}, it would not  be consistent with Nix and Hydra's small eccentricities and
inclinations.

Instead, we argue that Nix and Hydra formed within a collisional Plutocentric disk
that was composed of small bodies captured from heliocentric orbits.
It is quite plausible that at early times in the history of the Solar System there
were many small bodies in heliocentric orbit.
Three out of the four largest KBOs (including Pluto) have satellites
that have been suggested to form out of impacts
\citep{Brown06}.
 Charon formed out of an impact between
two Pluto-sized objects \citep{Canup}.
 Such
events are highly unlikely in the environment of the present day Kuiper
belt, where the time before the next impact between two Pluto-sized
bodies is $\sim 3\times 10^{13}$ yr.
This implies that in the past the velocity dispersion within the Kuiper belt
was much smaller than it is today, in which case gravitational focusing
would have enhanced the collision rate between Pluto-sized bodies.
The most likely
mechanism to cool the population of Plutos is dynamical friction with
a large mass of small bodies \citep{GLS}.  Today, these bodies have
suffered collisional diminuation and no direct evidence of their past
abundance remain. However, they must have been present, at least at some
point during the accretion growth period, to form the large KBOs in less
than the age of the Solar System \citep{KL98}.

The small bodies collide much more frequently and can be
collisionally accreted to the big bodies
\citep{SariGoldreich}. Collisions remove their random momenta but
cannot remove the net angular momentum. They then form a disk around
the big bodies.  Moons, formed or captured by such a disk, can be
migrated inward, be parked near resonant locations, and have small
free eccentricities (Shannon et al., in preparation).  The moons are
expected to be coplanar with Charon -- even if its nascent disk
started differently, such a dissipative disk can quickly relax to
Charon's orbital plane.

Hydra's orbit extends only to $\sim 1\%$ of Pluto's Hill radius.
Searches by \cite{Nicholson} and \cite{Steffl} have excluded the presence of
other comparably-sized moons in the Hill sphere. Why do present moons
occupy such a small fraction of the available space? We suspect the
moon-harboring disk may be limited in size, determined by the net
angular momentum of the accreted small bodies. 
Depending on the velocity dispersion of the small bodies in the 
circumsolar disk, the size of the accreted circum-Pluto disk might
be significantly smaller than Pluto's Hill radius.
Further exploration is
underway.

Pluto is not special.
Similar accretion disks may also have arisen around other large Kuiper belt objects. Do
they also possess small moons?  Hopefully the moons of Kuiper belt objects can teach
us about the early history of our Solar System, when the planets---and the KBO's themselves---were
being formed.

\appendix
\label{sec:appendix}

\section{A.  Hamiltonian Near the 4:1 Resonance}
\label{sec:hamapp}
Consider a massless test particle that is coplanar with the
Pluto-Charon binary, near its exterior 4:1 resonance.  Pluto and
Charon's mutual orbit has orbital elements $\left \{
a_C,e_C,\lambda_C,\pomega_C,n_C \right\}$, with
$\gn(M_P+M_C)=n_C^2a_C^3$ and $\lambda_C={\rm const}+n_Ct$.
We set $\pomega_C=0$ without loss of
generality. Since $M_C\ll M_P$, the effect of Charon on the particle
may be treated as a perturbation to the effect of Pluto.  The particle
has Pluto-centric orbital parameters\footnote{Throughout most of this paper, 
we use Jacobi co-ordinates, where the test particle's orbital elements
are relative to the barycenter of Pluto and Charon. 
 But 
 in this Appendix and in \S \ref{sec:destroy}, we employ Pluto-centric elements because
 this is traditional when using a disturbing function \citep{MD99}--even though 
 it  would be simple to use Jacobi elements instead \citep{LWdamp}.
 }
 $\{a,e,\lambda,\pomega\}$, and
its energy per unit mass is
\be
E=-{\gn M_P \over 2 a}-{\gn M_C \over a}{\cal R}
\label{eq:energy}
\ee
where ${\cal R}$ can be expanded as a sum of cosine terms. In Table
\ref{tab:dist} we list the coefficients and arguments of the leading
cosine terms near the 4:1 resonance.  To translate our
notation to that of Appendix B in
\cite{MD99}, from which we extracted the numerical constants in the
table, ${\cal R}={\cal R}_D+\alpha^{-2}{\cal R}_I$, where
$\alpha\equiv a_C/a$, and $\left\{B_1,B_2,C_0,C_1,C_2,C_3\right\}
=\left\{f_{10},f_2,f_{82},f_{83},f_{84},f_{85}-1/(3\alpha^2)\right\}$.
For the numerical constants, we set $\alpha=a_C/a_{\rm res}$, where
$a_{\rm res}$ is the semimajor axis at nominal 4:1 resonance
(eq. [\ref{eq:ares}]).

\begin{table}[t]
\centering
\caption{Leading cosine terms in ${\cal R}$
 (eq. [\ref{eq:energy}]) near 4:1 resonance, from Appendix B in
 \cite{MD99}}
\begin{tabular}{l|l|l}
\hline\hline
Cosine Argument & Coefficient & \\ \hline
$\pomega$ & $B_1e_Ce $ &  $B_1\equiv  -0.0902$ 
\\
0 & $B_2e^2$ & $B_2\equiv 0.0898$   
\\
$4\lambda-\lambda_C$ & $C_0e_C^3$ & $C_0\equiv  -0.285$ 
\\
$4\lambda-\lambda_C-\pomega$ & $C_1e_C^2e$ & $C_1\equiv 1.82$ 
\\
$4\lambda-\lambda_C-2\pomega$ & $C_2e_Ce^2$ & $C_2\equiv -3.82$ 
\\
$4\lambda-\lambda_C-3\pomega$ & $C_3e^3$ & $C_3\equiv 0.640$ \\ \hline
 \end{tabular}
\label{tab:dist}
\vspace{0.5cm}
\end{table}

 The Hamiltonian is equal to the energy, after replacing the
 particle's orbital elements with canonical variables
 $H(\Lambda,\lambda,\Gamma,\gamma)=E$.  We adopt Poincar\'e canonical
 variables $\{\Lambda,\lambda,\Gamma,\gamma \}$, where
\beqn
\Lambda&=&\left( \gn M_P a \right)^{1/2}  \label{eq:pa0}\\
\Gamma&=&\left(\gn M_P a\right)^{1/2}e^2/2 
\\
 \gamma&=&-\pomega \ ,
\eeqn
\citep{MD99}, dropping terms $O(e^4)$ from $\Gamma$.

To simplify the Hamiltonian, we employ a number of variable
transformations \citep{HM96,MD99}.  First, we absorb the
time-dependent parameter $\lambda_C=n_Ct+$const into $\lambda$ and
shift $\Lambda$ so that it vanishes at the nominal 4:1 resonance,
\be
\left\{\Lambda',\lambda'\right\}\equiv
\left\{
{\Lambda-\Lambda_{\rm res}\over 4}
,
4\lambda-
{\lambda_C}
\right\} \ ,
 \ee where $\Lambda_{\rm res}$ is a constant to be determined.  Since
 the generating function for this transformation is
 $F=(\Lambda'+\Lambda_{\rm res}/4)(4\lambda-\lambda_C)$, the new
 Hamiltonian is $H+\partial_t F$,
\be
H(\Lambda',\lambda';\Gamma,\gamma)=
-{(\gn M_P)^2\over 2
(4\Lambda'+\lambda_{\rm res})^2}-
(\Lambda'+{\Lambda_{\rm res}\over 4})n_C-
{\gn M_C\over a_{\rm res}}
{\cal R}
\label{eq:h1}
\ee
 setting $a$ to its value at nominal resonance, $a=a_{\rm res}$
 (eq. [\ref{eq:ares}]), everywhere except in the first two terms.To
 choose $\Lambda_{\rm res}$, we require that $\Lambda'=0$ at the
 nominal 4:1 resonance, which occurs where
 $0=(d/dt)(4\lambda-\lambda_C)=d\lambda'/dt= \partial H/\partial
 \Lambda'$. Since \be {\partial H\over \partial
 \Lambda'}\Big\vert_{\Lambda'=0}= {4\gn^2 M_P^2\over \Lambda_{\rm
 res}^{3}}-n_C \rightarrow 0\ , \ee we set \be \Lambda_{\rm res}\equiv
 (4\gn^2M_P^2/n_C)^{1/3} \ .  \ee The value of $a$ at nominal
 resonance is from equation (\ref{eq:pa0}),
\be
a_{\rm res}=4^{2/3}a_C 
\left(
{M_P\over M_C+M_P}
\right)^{1/3} \ . \label{eq:ares}
\ee
Expanding the Hamiltonian to second order in $\Lambda'$ and dropping
the constant term
\beqn
H(\Lambda',\lambda';\Gamma,\gamma)= -24{\gn M_P\over a_{\rm
res}}{\Lambda'^2\over \Lambda_{\rm res}^2}-{\gn M_C\over a_{\rm
res}}{\cal R}
\eeqn
We may rescale the momenta and Hamiltonian by the same constant factor
without altering the equations of motion.  Rescaling by $\Lambda_{\rm
res}$, the Hamiltonian becomes
\beqn
H(p_a,q_a;{e^2\over 2},-\pomega)&=&-{n_C\over 4}\left(24p_a^2+\mu {\cal R}\right) \\
\mu&\equiv& {M_C \over  M_P}
\eeqn
where 
\beqn
p_a&\equiv&{\Lambda'\over\Lambda_{\rm res}}\approx {1\over 8}{a-a_{\rm
res}\over a_{\rm res}} 
\label{eq:parelate}
\\
 q_a&\equiv&\lambda'\equiv 4\lambda-\lambda_C
 \label{eq:qarelate}
\eeqn 
The canonical momenta are $p_a$ and $e^2/2$, and their corresponding
conjugate co-ordinates are $q_a$ and $-\pomega$.

Next, we transform $\{e^2/2,-\pomega\}$ to remove the forced secular
eccentricity.  It simplifies the algebra to switch to the complex
canonical variable \citep{Strocchi66,Ogilvie07}
\be
Z\equiv ee^{-i\pomega} \ , \label{eq:zdef}
\ee
which is the usual complex eccentricity.
Hamilton's equations for $\{e^2/2,-\pomega\}$ are now expressed as
\be
{dZ\over dt} = 2i{\partial H\over\partial Z^*} \label{eq:zham}
\ee
where $Z^*$ is the complex conjugate of $Z$.  The secular part of
${\cal R}$ in Table \ref{tab:dist} is
\beqn
{\cal R}_{\rm sec}&=&e_CB_1{Z+Z^*\over 2} + B_2|Z|^2
\\
&=& B_2\left|Z+e_C{B_1\over 2B_2}\right|^2 \ ,
\eeqn
after dropping a constant. Therefore, we transform to the variable
\be
z\equiv Z + e_C{B_1\over 2 B_2} \ , \label{eq:ztrans}
\ee
which is the (complex) free eccentricity; the constant offset is the
forced eccentricity, $e_{\rm forced}=-e_C(B_1/2B_2)$.  Hamilton's
equation for $z$ is clearly the same as for $Z$ (eq. [\ref{eq:zham}]),
i.e., the transformation is canonical.

Under transformation (\ref{eq:ztrans}), the resonant part of ${\cal R}$ becomes
\beqn
{\cal R}_{\rm res}&=& {\rm Re}
\left(
e^{iq_a}\left(C_0e_C^3+C_1e_C^2Z+C_2e_CZ^2+C_3Z^3\right)
\right)
\\
&=&
 {\rm Re}
\left(
e^{iq_a}\left(c_0e_C^3+c_1e_C^2z+c_2e_Cz^2+c_3z^3\right)
\right) 
\label{eq:rres}
\eeqn
where, defining $\beta\equiv -B_1/2B_2$,
\beqn
c_0&=&C_0+C_1\beta+C_2\beta^2+C_3\beta^3=-0.26
\\
c_1&=&C_1+2C_2\beta+3C_3\beta^2 =-1.5
\\
c_2&=&C_2+3C_3\beta=-2.9
\\
c_3&=&C_3 = 0.64
\eeqn
Collecting results, the Hamiltonian is
\beqn
H(p_a,q_a;z)=-24p_a^2-\mu B_2|z|^2-  
\mu {\rm Re}
\left(
e^{iq_a}\left(c_0e_C^3+c_1e_C^2z+c_2e_Cz^2+c_3z^3\right)\right) 
\label{eq:hamfin}\ ,
\eeqn
after dropping the prefactor $n_C/4$, which means that time is now measured
in units of $4/n_C$.
Hamilton's equations of motion are
\be
\dot{p}_a={\partial H\over \partial q_a} \ 
;  \ \ 
\dot{q}_a=-{\partial H \over \partial p_a} \ 
;  \ \ 
\dot{z}=2i{\partial H\over \partial z^*} 
\label{eq:eom13}
\ee

\section{B. Tidal Dissipation}
\label{sec:tidal}

Assuming orbits and spins are coplanar, tidal dissipation due to
tides raised on Charon by Pluto are described by the following
evolution equations \citep{Hut}
\begin{eqnarray}
\left.{{da}\over{dt}}\right|_C & = & 
- {{6 k_{2C}}\over{T_C}} q (1+q) \left({{R_C}\over
a}\right)^8 {a \over{(1-e^2)^{15/2}}} \left\{ f_1-(1-e^2)^{3/2} f_2
{\Omega_C\over n}\right\},
\label{eq:dadtide}\\
\left.{{de}\over{dt}}\right|_C & = & 
- {{27 k_{2C}}\over{T_C}} q(1+q) \left({{R_C}\over
a}\right)^8 {e \over{(1-e^2)^{13/2}}} \left\{ f_3-{{11}\over{18}}
(1-e^2)^{3/2} f_4 {\Omega_C\over n}\right\},
\label{eq:dedtide}\\
{{d\Omega_C}\over{dt}} & = &  {{3 k_{2C}}\over{T_C}} {{q^2}\over{r_{gC}^2}} 
{n\over{(1-e^2)^6}} \left({{R_C}\over
a}\right)^6 \left\{ f_2-
(1-e^2)^{3/2} f_5 {\Omega_C\over n}\right\},
\label{eq:domegadt}
\end{eqnarray}
where $q = M_P/M_C$ is the mass ratio, 
$\Omega_C$ is Charon's spin rate, $n = \sqrt{G (M_P+M_C)/a^3}$
is the orbital mean motion, $k_{2C}$ is the tidal Love number of
Charon, and $r_{gC}$ its radius of gyration related to the moment of
inertia by $I = r_g^2 m R^2$, with $r_g \sim 0.6$ for a uniform density
sphere.  The values for $f_i$ are all of order unity, $f_i = 1 +
{\cal{O}}(e^2) + {\cal{O}}(e^4)...$, with their exact expressions
listed in \citet{Hut}.  In this tidal model, $T_C = {{R_C^3}/{G M_C \tau_C}}$
where $\tau_C$ is the (assumed) constant tidal lag time. To connect this
model to that of \citet{GoldreichSoter} which assumes a constant lag phase (and
a constant tidal $Q$ factor), we let the tidal lag time $\tau_C = 1/(2 n Q_C )$, 
with $Q_C$ being Charon's tidal dissipation quality factor. This choice
is somewhat arbitrary: one may also assume, e.g., $\tau_C = 1/[2 (\Omega_C-n) Q_C]$.
The tide raised by Charon on Pluto causes similar effects
but with all subscripts $C$ substituted by $P$ and the mass ratio inverted,
$q = M_C/M_P$.  The net orbital evolution is given by the sum of both tides.

The Love number for a uniform density sphere is \citep[see,
e.g.,][]{Dob97,MD99}
\be
k_2 = {{3/2}\over{1 + {\tilde \mu}}},
\label{eq:k2}
\ee
where ${\tilde \mu} = 19 \mu/(2 \rho g R)$ is the effective rigidity of
the body,  $\mu$ the material strength, $\rho$ the density
and $g$ the gravitational acceleration. If the body is a rubble pile, ${\tilde \mu} \ll
1$ and $k_2 \sim 3/2$. However, if the body is a consolidated ice
sphere with $\mu \sim 4\times 10^{10} {\rm dyne}/\cm^2$,
$k_{2C} \sim 0.005$ and $k_{2P} \sim 0.05$ \citep{Dob97}.

The following are numerical estimates for the various timescales,
 $\tau_x \equiv x/|\dot x|$,
\begin{eqnarray}
\left. \tau_a\right|_C & \sim &  
10^8 {\rm yrs} \left({Q_C}\over{100}\right) 
\left({k_{2C}}\over{0.005}\right)^{-1} 
\left({{a_c}\over{17 R_P}}\right)^{6.5} ,
\label{eq:time1}\\
\left. \tau_a\right|_P & \sim &
2\times 10^7 {\rm yrs} \left({Q_P}\over{100}\right) 
\left({k_{2P}}\over{0.05}\right)^{-1} 
\left({{a_c}\over{17 R_P}}\right)^{6.5} ,
\label{eq:time2}\\
\left. \tau_e\right|_C & \sim &
3\times 10^7 {\rm yrs} \left({Q_C}\over{100}\right) 
\left({k_{2C}}\over{0.005}\right)^{-1} 
\left({{a_c}\over{17 R_P}}\right)^{6.5},\label{eq:time3}\\
\left. \tau_e\right|_P & \sim &
5\times 10^6 {\rm yrs} \left({Q_P}\over{100}\right) 
\left({k_{2P}}\over{0.05}\right)^{-1} 
\left({{a_c}\over{17 R_P}}\right)^{6.5},\label{eq:time4}\\
\tau_{\Omega_C} & \sim &
6\times 10^4 {\rm yrs} \left({Q_C}\over{100}\right) 
\left({k_{2C}}\over{0.005}\right)^{-1} 
\left({{a_c}\over{17 R_P}}\right)^{4.5},\label{eq:time5}\\
\tau_{\Omega_P} & \sim &
6\times 10^5 {\rm yrs} \left({Q_P}\over{100}\right) 
\left({k_{2P}}\over{0.05}\right)^{-1} 
\left({{a_c}\over{17 R_P}}\right)^{4.5},
\label{eq:time6}
\end{eqnarray}
where we have scaled $Q_P$ and $Q_C$ by values typical of solid bodies
and have taken the limit $e \ll 1$. 

Charon's spin is quickly (pseudo-)synchronized with the orbit, and 
Pluto's spin synchronization is on-going
until the system reaches the final state of double synchronization.
So most of the $a$ evolution is contributed by tides on Pluto (eq. \ref{eq:time2}).

During the orbital expansion, the tide on Pluto increases $e$ while the tide on Charon decreases it.
With our choice of parameters, the former dominates over the latter. Once Charon
reaches its equilibrium location, both tides damp the eccentricity.

If the spin direction of body is misaligned with the orbit normal, it
will be tilted to alignment in roughly the spin synchronization time \citep{Hut}.

\bibliographystyle{apj}
\bibliography{plutoy}

\end{document}